# Superconductivity in nickel-based compound GdONiBi and hole doped $Gd_{0.9}Sr_{0.1}ONiBi$

Junyi Ge, Shixun Cao, Jincang Zhang

Department of Physics, Shanghai University, Shanghai 200444, China

We successfully synthesized the nickel-based compound GdONiBi with superconducting transition temperature about 4.5 K. By partially substituting the element Gd with Sr to introduce holes into the material, we got new superconductor $Gd_{0.9}Sr_{0.1}ONiBi$ with critical temperature about 4.7 K. The normal state resistivity in nickel-based samples shows a metallic behavior. The magnetoresistance measurements show a different behavior compared to those in iron-based compounds which indicates that the mechanism in the two kinds of superconductors maybe different.

## Ⅰ. Introduction

The work of searching for new kind of superconductors has never been stopped. Recently the discovery of the novel superconductor with the $T_c$ up to 55 K in layered oxypnictides ROTmPn (R=La, Ce, Sm, Tm=Fe, Co, Ni, Pn=P, As, Bi)[1,2,3,4,5,6,7,8,9,10] has attracted great interest in this system since it is the first class of material which exhibits superconductivity at high temperatures without copper. As it contains magnetic elements Tm (Fe, Ni) and some special properties we have known [11], the original of superconductivity in this kind of material maybe quite different from those superconductors discovered before such as copper-oxides, $MgB_2$ and so on. Therefore, it provides us potential possibilities searching for novel superconductors with higher critical temperature as well as the key to the mechanism of high $T_c$ superconductors.

Yet until now most of the research groups are focusing on the iron-based superconductors of arsenide. And LnOFeAs can only become superconductive when doped with hole or electric carriers while LaONiBi itself is reported to be superconducting [12]. In order to further studying the mechanism and the relationship between arsenides and bismuthides, here for the first time we report the newly synthesis and studies of another two superconductors based on nickel GdONiBi and hole doped superconductor $Gd_{0.9}Sr_{0.1}ONiBi$.

## II. Experimental

The sample was prepared with two step method. First, GdBi alloy was prepared by arc melting appropriate quantities of gadolinium metal (99.9%, Alfa) and bismuth (99.99%) with a slight excess (5~10 wt.%) of bismuth on a water-cooled Cu crucible under high purity argon gas atmosphere. Then the resultant precursor was thoroughly grounded together with Ni powder (99.5%), $Ni_2O_3$ powder (99.99%), and $SrCO_3$ (99.99%) in stoichiometry of $Gd_{1-x}Sr_xONiBi$ ($x$=0, 0.1). The mixtures were pressed into pellets and sealed in a quartz tube with high vacuum and then annealed at 850 $^oC$ for 36 hours. The measurements were carried out on a physical property measurement system (PPMS, Quantum Design) with magnetic fields up to 9 T. All the transport measurements were employed by a four-probe method employing silver paste as electrodes.

## III. Results and discussion

As is known that in rare earth nickel borocarbide $ErNi_2B_2C$, the superconductivity of $ErNi_2B_2C$ could be strongly depressed by the substituting of Er with Gd, which is ascribed to the magnetic moment of gadolinium element [13]. In Fig.1(a) we present the resistive transport of GdONiBi and $Gd_{0.9}Sr_{0.1}ONiBi$ in zero field. It is seen clearly that the resistivity experience a sharp drop to zero at 4.5 K with the transition width about 0.8 K for GdONiBi while for the Sr-doped sample, the transition temperature is slightly increased to 4.7 K. It seems that the moment of rear earth metal has little effect on the superconductivity of LnONiBi (in LaONiBi, $T_c$ is about 4.25 K). The hole doped process also increases the magnitude of the normal state resistivity. Both samples show metallic behavior in the whole range above transition temperature which is different from those in iron based superconductors like $LaO_{1-x}F_xFeAs$ whose resistivity exhibiting an anomaly peak at ~ 150 K. The critical temperature decreases as the applied field increases as shown in the inset of Fig.1(b). The upper critical field $H_{c2}$ shows approximately linear $T$-dependence with the slope dHc2/dT≈0.8 T/K which is very close to that in LaONiBi[12]. The DC susceptibility data in Fig.2 drops to diamagnetic signal at 4.2 K which also confirms the superconductive transition. Yet there are still some impurities in the sample such as GdNi and NiBi which make it shown weak ferromagnetic behavior at high temperatures as shown in the inset of Fig.2. Hence, further study of synthesizing materials with more superconducting phase is still needed.

Theoretical calculations as well as experiments already reveal that the iron based superconductors exhibit multi-band features[14, 15, 16, 17, 18, 19]. According to Kozhevnikov's report, LaONiBi also has a typical feature of two-band metal[12]. As is known that for a single-band free-electron system, there is no magnetoresistance (MR), while in a system with two or more bands there will be magnetoresistance such as in $MgB_2$[19]. In order to get a deep insight of characteristic of the nickel based bismuthide, we studied the MR effect of GdONiBi and $Gd_{0.9}Sr_{0.1}ONiBi$. In this paper the magnetoresistivity was defined by $\Delta\rho = \rho(H)-\rho_0$, where $\rho(H)$ is the resistivity under applied magnetic field $H$ and $\rho_0$ is the zero field resistivity. Fig.3(a) shows the MR of GdONiBi at different temperatures from 20 K to 300 K with the magnetic field up to 9 T. It can be seen that the MR is very small with the maximum about 1.3%. The magnitude of the MR became smaller with increasing temperature which is very similar with that in iron based superconductors[20]. A method to reveal if a system has a multi-band effect or not is through Kohler's law. For a single-band system, the slope of $\Delta\rho/\rho_0$ vs $H/\rho_0$ should be a straight line which is independent of temperatures[21]. The inset of Fig3(a) is the Kohler plot of GdONiBi. As is seen the Kohler's Law is not obeyed with the slope of MR vs $H/\rho_0$ changes at different temperatures. The result may indicate a two-band or multi-band structure in the nickel based superconductors. The MR of hole doped $Gd_{0.9}Sr_{0.1}ONiBi$ has almost the same behavior as GdONiBi (not shown) except for the magnitude about two times bigger than that of GdONiBi which may indicate that the hole doped compound has a more complex band structure. Fig.3(b) shows the temperature dependence of MR at 9 T for the two samples. Also we found an abnormality compared with iron based compounds [19]. In GdONiBi with temperature increases, the MR first dropped to a constant value of 0.4%, and then experienced another drop at 125 K, while in $Gd_{0.9}Sr_{0.1}ONiBi$ the MR achieves a peak value in 125K.

## Ⅳ. Conclusion

In summary, for the first time we have synthesized nickel based superconductor bismuthide GdONiBi, and by substituting Gd with Sr, the transition temperature could be increased to 4.7 K. We studied the magnetoresistance in both samples, and found some different properties compared with iron based arsenides which may indicate that the mechanism of superconductivity in bismuthides is different from the arsenides. Also we found a multi-band like structure in this superconductor. For further understanding the mechanism of superconductivity in this novel class of superconductors, the study in bismuthide superconductors may provide us another way to get the point.

## Ⅴ. Acknowledgement

This work is supported by the National Natural Science Foundation of China (Grant Nos.10674092, 10774097), and the Shanghai Leading Academic Discipline Project (Grant No.T0104).

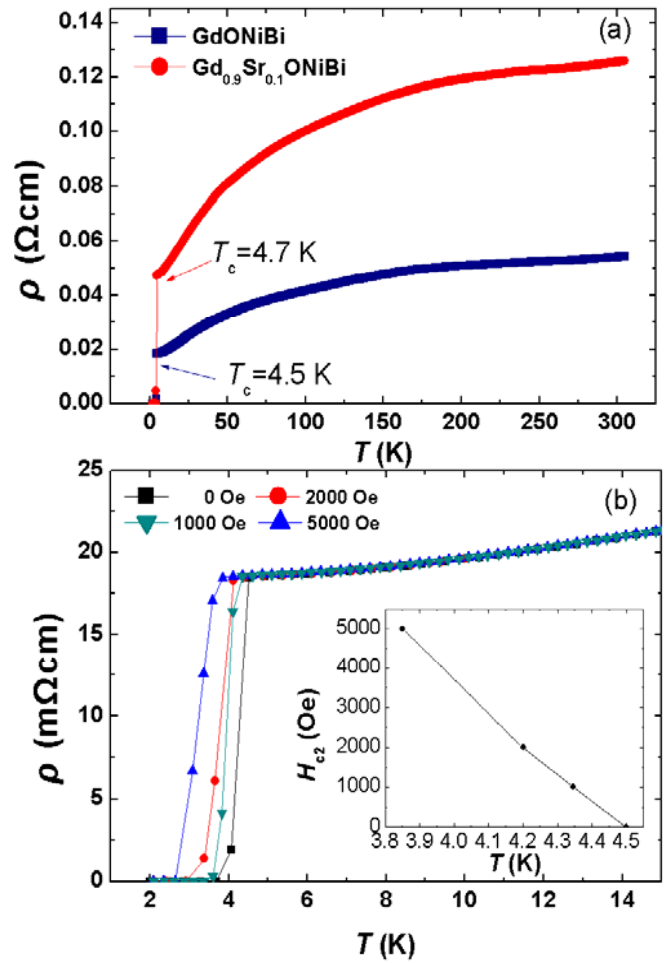

Fig.1 (a) Temperature dependence of resistivity for sample GdONiBi and hole doped Gd0.9Sr0.1ONiBi. (b) The temperature dependence of resistivity of the sample GdONiBi under different magnetic fields. It's clear that the transition temperature decreases with applied field increases. The inset shows the linear dependence of the upper field with temperature.

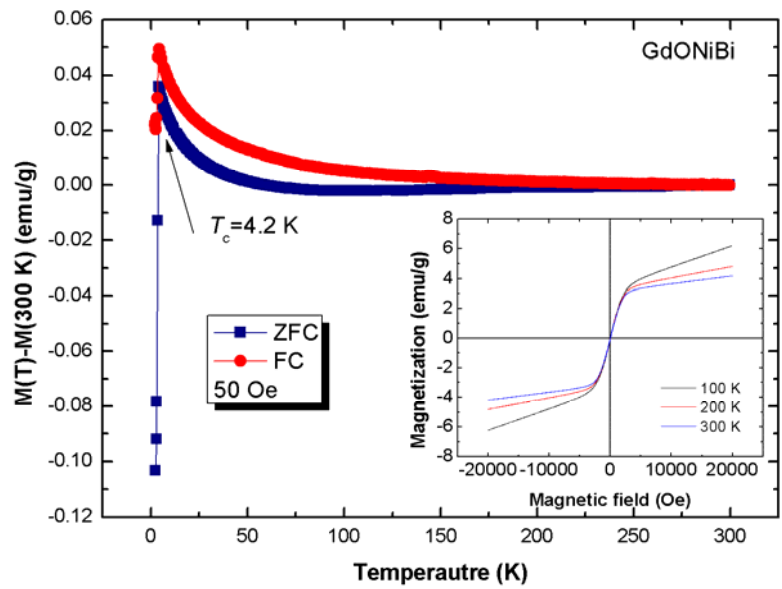

Fig.2 DC magnetization of GdONiBi at 50 Oe measured in the zero-field-cooled (ZFC) and field-cooled (FC) processes. The inset shows weak ferromagnetic behavior at high temperatures which indicate some ferromagnetic impurities in the sample.

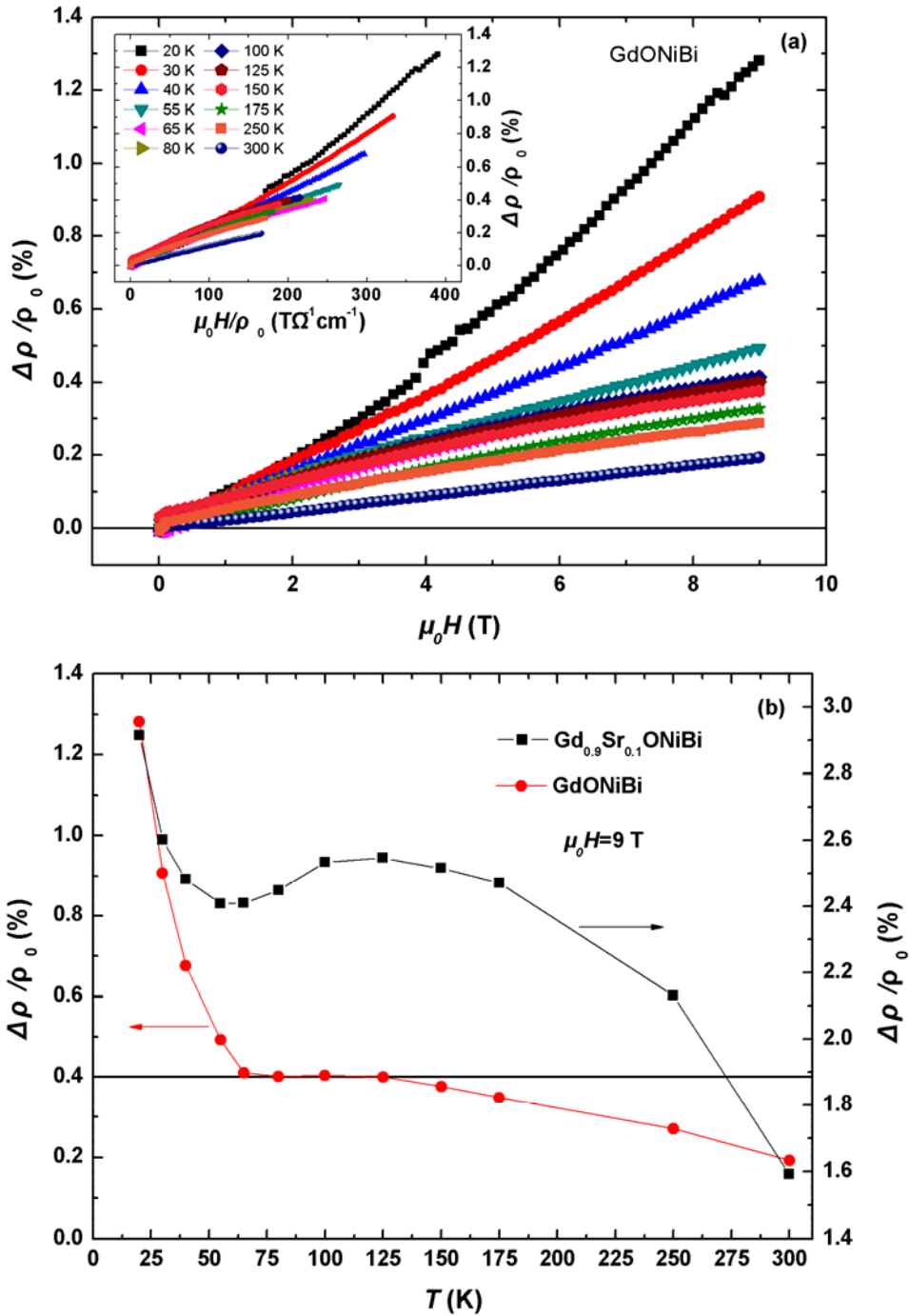

Fig.3 (a) Magnetic field dependence of magnetoresistance at different temperatures for the sample GdONiBi. The inset shows the Kohler's plot for magnetoresistance at different temperatures. (b) Temperature dependence of magnetoresistance for bismuthide GdONiBi and hole doped $Gd_{0.9}Sr_{0.1}ONiBi$.